\documentclass[12pt]{article} 

\usepackage{latexsym} 	
\usepackage{amssymb}  	
\usepackage{amsbsy}
\usepackage{epsfig}     
\usepackage{graphics,color}
\usepackage{a4}

\newcommand{\lpc}{{g^2}}
\newcommand{\om}{\omega}
\newcommand{\Om}{\Omega}

\newcommand{\Tr}{\mbox{Tr}}

\newcommand{\C}{{\cal C}}
\newcommand{\bra}{\langle}
\newcommand{\ket}{\rangle}

\newcommand{\half}{\frac{1}{2}}

\newcommand{\veck}{{\mathbf k}}

\newcommand{\vecx}{{\mathbf x}}

\newcommand{\vecnul}{{\mathbf 0}}  
\newcommand{\kv}{{\mathbf k}}

\newcommand{\xv}{{\mathbf x}}
\newcommand{\yv}{{\mathbf y}}

\newcommand{\be}{\begin{equation}}
\newcommand{\ee}{\end{equation}}
\newcommand{\ba}{\begin{eqnarray}}
\newcommand{\ea}{\end{eqnarray}}
\newcommand{\bea}{\begin{eqnarray}}
\newcommand{\eea}{\end{eqnarray}}
\newcommand{\bean}{\begin{eqnarray*}}
\newcommand{\eean}{\end{eqnarray*}}
\newcommand{\nn}{\nonumber}
\newcommand{\hm}{\hspace*{-0.6cm}}

\renewcommand{\theequation}{\arabic{section}.\arabic{equation}}

\begin{document}

\title{
\vskip -100pt
{\begin{normalsize}
%\mbox{} \hfill DAMTP-2007-???\\
\mbox{} \hfill arXiv:0712.1120 [hep-ph] \\
\vskip  100pt
\end{normalsize}}
{\bf\Large Thermal effects on slow-roll 
dynamics}
\author{
\addtocounter{footnote}{2}
Gert Aarts$^{a}$\thanks{email: g.aarts@swan.ac.uk}
 {} and
Anders Tranberg$^{b,c}$\thanks{email: anders.tranberg@oulu.fi}
 \\ {} \\
{}$^a${\em\normalsize Department of Physics, Swansea University}
\\
{\em\normalsize Singleton Park, Swansea SA2 8PP, United Kingdom}
\\ {} \\
 {}$^b${\em\normalsize DAMTP, University of Cambridge} \\
   {\em\normalsize Wilberforce Road, Cambridge CB3 0WA, United Kingdom}
\\ {} \\
\addtocounter{footnote}{-3}
 {}$^c${\em\normalsize Department of Physics, University of Oulu} \\
   {\em\normalsize P.O.\ Box 3000, FI-90014 Oulu, 
   Finland\footnote{Present address}}
}
}
\date{December 7, 2007}
\maketitle
\begin{abstract}
 A description of the transition from the inflationary epoch to radiation 
domination requires the understanding of quantum fields out of thermal 
equilibrium, particle creation and thermalisation. This can be studied 
from first principles by solving a set of truncated real-time 
Schwinger-Dyson equations, written in terms of the mean field (inflaton) 
and the field propagators, derived from the two-particle irreducible 
effective action.
 We investigate some aspects of this problem by considering the dynamics 
of a slow-rolling mean field coupled to a second quantum field, using a 
$\varphi^{2}\chi^2$ interaction. We focus on thermal effects. It is found 
that interactions lead to an earlier end of slow-roll and that the 
evolution
 afterwards depends on details of the heatbath.
 \end{abstract}
                                                                                
\newpage
                                                                                
%%%%%%%%%%%%%%%%%%%%%%%%%%%%%%%%%%%%%%%%%%%%%%%%%%%%%%%%%%%%%%%%%%%%%
                                                                                
% SECTION INTRODUCTION           
                                                                     
\section{Introduction}
\label{sec:Introduction}
\setcounter{equation}{0}

Cosmological observations strongly suggest that a stage of accelerated
expansion took place in the very early Universe
\cite{Spergel:2006hy,Peiris:2003ff}. Such a stage is well described in
terms of the dynamics of a scalar field, the inflaton $\phi(t)=\bra
\varphi(t,\xv) \ket$, slow-rolling in a suitable potential $V[\phi]$.
Classically, the inflaton equation of motion reads
 \be 
 \label{eq:inf1}
 \ddot\phi(t) + 3H(t)\dot\phi(t) + V'[\phi(t)] = 0, 
\ee 
with the Hubble rate determined by the Friedmann equation 
\be
\label{eq:friedmann}
H^2(t)=\frac{\left(\half \dot\phi^2(t)+V[\phi(t)]\right)}{3M_{\rm pl}^2},
\ee
where $M_{\rm pl}$ is the Planck mass. 

 It is natural to assume that the inflaton is coupled directly or 
indirectly to the Standard Model fields and that after inflation, the 
energy stored in the inflaton is transfered to excitations of these 
fields. This can be realised through perturbative reheating 
\cite{Dolgov:1989us,Kolb:1990vq} or nonperturbative preheating 
\cite{Traschen:1990sw,Kofman:1997yn,Khlebnikov:1996mc,Prokopec:1996rr, 
Boyanovsky:1995em,Boyanovsky:1996sq,Felder:2000hj,GarciaBellido:2002aj,Arrizabalaga:2004iw} 
processes. Through scattering, the system eventually reaches thermal 
equilibrium and a radiation dominated Universe emerges.
 In quantum field theory, the inflaton evolution in real time is described 
by a Schwinger-Dyson equation,
 \be
 \label{eq:inf3}
 \ddot\phi(t) + 3H(t)\dot\phi(t) + V'[\phi(t)]
 = -\int_{t_0}^t dt'\, \Sigma_{\phi}(t,t')\phi(t'),
 \ee
 where $t_0=0$ is the initial time, taken as the starting point for 
the evolution.  The nonlocal (self-energy like) term $\Sigma_{\phi}(t,t')$ 
contains the interaction with quantum fields and is to be determined, 
either through some well-motivated ansatz or in terms of a systematic 
diagrammatic expansion.

Due to the interactions, one may expect a backreaction on the 
inflaton resulting in, e.g., a modified effective mass parameter and 
possibly friction. A phenomenological inflaton equation of motion 
incorporating this would be
 \be 
 \label{eq:inf2}
 \ddot\phi(t) + \left[ 3H(t) + \Upsilon(t) \right] \dot\phi(t) 
 + V_{\rm eff}'[\phi(t)] = 0. 
 \ee 
 Various approximations have been invoked to motivate Eq.\ (\ref{eq:inf2}) 
\cite{Hosoya:1983ke,Moss:1985wn,Morikawa:1986rp,Berera:2001gs,Berera:2007qm}.\footnote{See 
also Refs.\ 
\cite{Boyanovsky:1994me,Lawrie:2002wm,Lawrie:2002zd} for 
objections to this equation.} Effective inflaton dynamics as described by 
Eq.\ (\ref{eq:inf2}) is important for warm inflation 
\cite{Moss:1985wn,Berera:1995ie}, in which it is assumed that particles 
created during inflation 
  interact fast enough such that a quasi-stable nonvacuum state is 
reached.

 For a mean field $\phi(t)$ undergoing small amplitude oscillations in a
thermal background, an effective damping rate $\Upsilon$ is indeed
generated, and it can be calculated perturbatively in a linear response
treatment \cite{Arrizabalaga:2005tf}. However, when the mean field
displacement from equilibrium cannot be treated as a small perturbation,
as in the case of a slow-rolling inflaton, things are less clear and a
fully nonequilibrium treatment is required.  This is even more the case
if properties of the heatbath depend on the value of the mean field
through interactions.

 In this paper we will not attempt a full investigation in an expanding 
background. Instead, we will study some aspects of this problem in a 
simple inflation-inspired model using the tools of nonequilibrium field 
theory, i.e.\ solving the Schwinger-Dyson equation (\ref{eq:inf3}), as 
derived from truncations of the two-particle irreducible (2PI) effective 
action \cite{Cornwall:1974vz,Berges:2004yj}.
 In section \ref{sec:model} we introduce the model, explain how we deal 
with the expansion and describe the dynamics of the mean field and 
fluctuations in absence of most interactions. We write down the 
Schwinger-Dyson equations to next-to-leading order in a coupling expansion 
of the 2PI effective action in Section \ref{sec:2PI}.
  In Section \ref{sec:numerics} we subsequently solve the lattice 
discretised integro-differential equations numerically without further 
approximations. We consider various scenarios and study in particular the 
role of thermal initial conditions for the two quantum fields in our 
model. Our findings are summarized in Section \ref{sec:conclusions}. 
 In the Appendix we collect several approximations that are well-known in 
the literature and that are used in the main body of the paper for 
comparison.

% SECTION: MODEL

 \section{Model and free evolution }
 \label{sec:model}
 \setcounter{equation}{0}

 We consider two interacting scalar fields $\varphi$ and $\chi$ evolving 
in a flat Friedmann-Robertson-Walker Universe, with the action 
 \be
 \label{eq:action} 
 S = -\int d^4x \,\sqrt{-g} 
 \left[ \frac{1}{2}g^{\mu\nu}
\partial_\mu\varphi \partial_\nu\varphi +
\frac{1}{2}g^{\mu\nu}\partial_\mu \chi \partial_\nu\chi
+\frac{1}{2}m_{\varphi}^{2}\varphi^{2}+\frac{1}{2}m_{\chi}^{2}\chi^{2}
+ \frac{1}{2}g^2\varphi^{2}\chi^{2}\right].
 \ee 
The metric is specified by $ds^2=dt^2-a^2(t)d\xv^2$. Hence, the 
spatial derivatives are given by 
$\partial_{\bf x}/a(t)$, where ${\bf x}$ is co-moving. The Friedmann 
equation determines the evolution of the scale factor as
 \bea
 \left(\frac{\dot{a}(t)}{a(t)}\right)^{2} \equiv H^{2}(t) = \frac{\langle
T^{00}(t)\rangle}{3M^{2}_{\rm pl}},
 \eea
 where $\bra T^{00}\ket$  is the renormalized expectation value of the 
energy density of matter. 
 We let $\varphi$ play the role of the inflaton: its expectation value 
$\langle\varphi(t,{\bf x})\rangle =\phi(t)$ is homogeneous and set to some 
large value initially, $\phi(0)=\phi_{0}$, prompting slow-roll inflation.  
We take $\langle\chi(t,{\bf x})\rangle=0$ for all times, consistent with 
the symmetries of the action.

 In order to understand the dynamics in this model, we start by 
considering the classical evolution of $\phi(t)$ without interactions.
Its equation of motion is
 \be
 \label{eq:mf1}
 \ddot{\phi}(t)+3H(t)\dot{\phi}(t)+m_{\varphi}^{2}\phi(t)=0,
 \ee
 with $H$ given by
 \be
 \label{eq:H}
 H^{2}(t)=\frac{\dot{\phi}^{2}(t)+m_{\varphi}^{2}\phi^{2}(t)}{6M_{\rm pl}^2}.
 \ee
 In the slow-roll limit, $H^{2}(t) = m_{\varphi}^{2}\phi^{2}(t)/6M^{2}_{\rm
pl}$, and we find
 \be
 \label{eq:SRphi}
 \phi(t)=\phi_0 - \sqrt{\frac{2}{3}}m_{\varphi}M_{\rm pl}t.
 \ee
 Assuming that inflation ends when the slow-roll parameter $\epsilon = 
\half M_{\rm pl}^2 (V'[\phi]/V[\phi])^2$ equals 1, we find that there is 
inflation when $\phi/M_{\rm pl} \gtrsim \sqrt{2}$, irrespective of the 
value of $m_\varphi$; a given initial value $\phi_{0}$ corresponds in the 
slow-roll approximaton to $N_{e}=[ (\phi_0/M_{\rm pl})^2-2]/4$ e-folds 
before the end of inflation.

 The $\chi$ field, coupled to this slow-rolling inflaton, is expanded 
 in terms of creation and annihilation operators as
 \be
 \chi(t,{\bf x}) = \int_\kv \left(a_{\bf k}f_{\bf k}(t)
 e^{i{\kv\cdot\xv}}+a^{\dagger}_{\bf k}f^{*}_{\bf k}(t)e^{-i{\kv\cdot\xv}}
 \right),
 \ee
where
\be
 \int_\kv = \int \frac{d^3k}{(2\pi)^3}.
\ee
 The mode functions $f_{\bf k}(t)$ obey the equation
 \be
 \label{eq:modeeq}
 \left[\partial_{t}^{2}+3H(t)\partial_{t}+{\bf
k}^{2}/a^{2}(t)+m_{\chi}^{2}+\lpc\phi^{2}(t)\right]f_{\bf k}(t)=0.
 \ee
 At this level the inflaton acts as a time-dependent mass.

 We now make a practical choice that we will follow throughout this paper: 
we ignore the expansion of the Universe for the quantum fields (or mode 
functions) and put $H=0$, $a(t)=1$ in Eq.\ (\ref{eq:modeeq}), while 
keeping Hubble friction in the mean field equation. This has the effect of 
emphasizing the role of scattering and energy transfer during the 
slow-roll regime and maximises the backreaction of the quantum modes on 
the mean field by omitting the effect of dilution and redshift. As a 
result we are not studying inflation, but rather the impact of 
interactions on a slow-rolling mean field. It allows us to disentangle the 
direct effect of field theory interactions from the effect of including a 
$\chi$-component in $T^{00}$. Such a component would influence the 
inflaton evolution via the Friedmann equation.
 Neglecting expansion is expected to strongly favour thermalisation. We 
can make a rough estimate by comparing the expansion rate $H$ to the 
perturbative damping rate $\Gamma(T)$ \cite{Arrizabalaga:2005tf}. The 
ratio $\Gamma/H$ is maximal at the end of inflation, and for the largest 
temperatures used in Section \ref{sec:numerics} ($T/M_{\rm pl}=4$), we 
find $\Gamma/H\sim 0.03$. The expansion could therefore have a significant 
impact on thermalisation. In practice however, the time scales under 
consideration here are much shorter than the thermalisation time scale, 
and, as we will see below, equilibration and thermalization will not play 
a role here.

 \begin{figure}[t]
 \begin{center}
 \epsfig{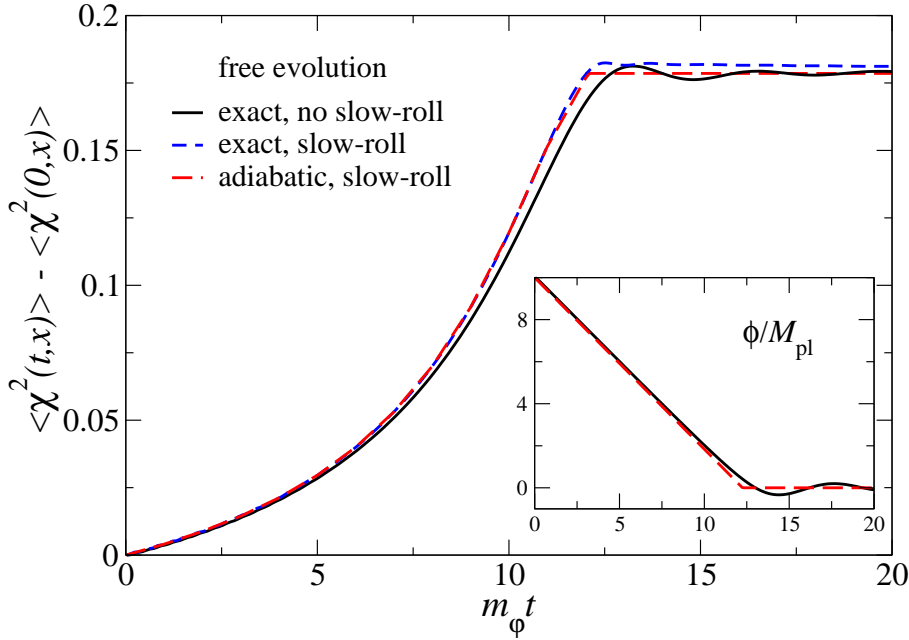}
 \caption{The equal-time correlator 
$\bra\chi^2(t,\xv)\ket-\bra\chi^2(0,\xv)\ket$ for free evolution of the 
$\chi$ modes in the mean-field background. Shown are the numerically 
determined evolution in the exact (full) and the slow-roll (dashed) 
background, as well as in the adiabatic approximation (long-dashed).
 The inset shows the mean field during the exact evolution (full) and in 
the slow-roll approximation (dashed).
 }
 \label{fig:tads}
 \end{center}
 \end{figure}

 In Fig.\ \ref{fig:tads} we show the evolution of the mean field $\phi(t)$ 
and the (subtracted) equal-time correlator
\be
 \bra \chi^2(t,\vecx) \ket  
= \int_\kv  \left| f_\veck(t) \right|^2,
\ee
for an initial state in vacuum.
 The initial value of the mean field is $\phi_0/M_{\rm pl}=10>\sqrt{2}$, 
ensuring an extended slow-roll stage. 
 We show the 
 evolution of $\bra\chi^2\ket$ determined by numerically solving the 
equations for the mode functions $f_\kv(t)$ using various approximations 
detailed in Appendices \ref{sec:WKB} and \ref{sec:exact}: in the exact 
mean-field background determined by Eqs.\ (\ref{eq:mf1},\ref{eq:H}), in 
the slow-roll background (\ref{eq:SRphi}), and in the adiabatic 
approximation.
 In the inset the time evolution of the mean field is shown, determined by 
solving Eqs.\ (\ref{eq:mf1},\ref{eq:H}) numerically and in the slow-roll 
approximation (\ref{eq:SRphi}). In the latter the evolution is stopped 
when 
$\phi(t)$ reaches zero; $\phi(t)$ is kept constant at zero from then on.
 
 During the slow-roll regime the mean field $\phi(t)$ and hence the 
effective $\chi$ mode energies 
$\omega_\kv(t)=[\kv^2+m_\chi^2+g^2\phi^2(t)]^{1/2}$ decrease. As a 
result 
the equal-time correlator $\bra\chi^2\ket$ increases in time. The sharp 
increase is described very well in the adiabatic approximation. Therefore 
it does not correspond to a large amount of particle production.
 The question we address in the remainder of the paper is whether and how 
interactions modify this basic scenario.

% SECTION: 2PI EXPANSION AT WEAK COUPLING

\section{2PI expansion at weak coupling\label{sec:2PI}}
 \setcounter{equation}{0}

Corrections to the dynamics considered above arise from 
quantum and possibly thermal back-reaction effects. These can be deduced 
from taking the expectation value of the Heisenberg operator equation of 
motion, yielding
 \be
 \label{eq:backreact0}
 \partial_t^2 \langle \varphi(x)\rangle 
 + 3H(t) \partial_t \langle\varphi(x)\rangle 
 + m_\varphi^2 \langle\varphi(x)\rangle 
 + \lpc \langle \chi^2(x)\varphi(x)\rangle = 0.
 \ee
 The last term can be expanded in terms of connected correlators,
 \be
\label{eq:heisexp}
 \langle\chi^{2}(x) \varphi(x)\rangle = 
 \langle\chi^{2}(x)\rangle\phi(t) 
 + \langle\chi^2(x)\varphi(x)\rangle_{\rm connected},
 \ee
using again that $\bra\chi(x)\ket=0$
and writing $\langle\varphi(x)\rangle = \phi(t)$. 
As is well-known, keeping only the first term amounts to the Hartree or 
Gaussian approximation (see Appendix \ref{sec:hartree}). In that case the 
effects of interactions reduce to a time-dependent mass.
In order to systematically go beyond the Hartree approximation and include 
effects from the connected three-point correlator in Eq.\ (\ref{eq:heisexp}), 
we use a formalism that allows for consistent truncations of the 
Schwinger-Dyson hierarchy: the two-particle irreducible (2PI) effective 
action.

 The 2PI effective action is a functional written in terms of the mean 
field $\phi$ and the propagators $G_{\varphi}$, $G_{\chi}$ 
\cite{Cornwall:1974vz},
 \bea
 \Gamma[\phi,G_{\varphi},G_{\chi}] = &&\hm S[\phi] + \frac{i}{2}\Tr\ln\,
G_{\varphi,0}^{-1}\, + \frac{i}{2}\Tr\, G_{\varphi,0}^{-1}G_{\varphi,0}
 \nonumber\\
 &&\hm + \frac{i}{2}\Tr\ln\,G_{\chi,0}^{-1}\, + \frac{i}{2}\Tr\,
G_{\chi,0}^{-1}G_{\chi,0} +
\Gamma_{2}\left[\phi,G_{\varphi},G_{\chi}\right],
 \eea
 where $S[\phi]$ is the tree-level action (\ref{eq:action}) written in 
terms of the mean field, $G_{\varphi/\chi,0}^{-1}$ are the free inverse 
propagators,
 \bea
 iG_{0,\varphi}^{-1}(x,y) = &&\hm -\left[ \square_x + m_\varphi^2 \right]
 \delta_{\mathcal{C}}(x-y),
 \\
 iG_{0,\chi}^{-1}(x,y) = &&\hm -\left[ \square_x + m_\chi^2 + 
 \lpc\phi^2(x) \right] \delta_{\mathcal{C}}(x-y),
 \eea
 and $\Gamma_{2}$ is the sum of all two-particle-irreducible diagrams, 
depending on full propagators and the mean field $\phi$ via the 
three-point vertex $g^2\phi\varphi\chi^2$. We follow the notation of Ref.\ 
\cite{Aarts:2002dj}. 

\begin{figure}
 \begin{center}
 \epsfig{file=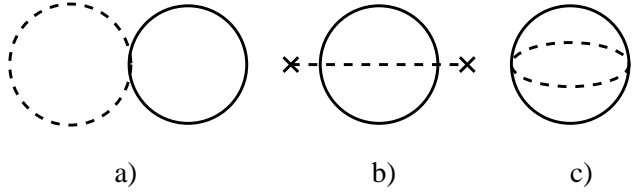,height=2.5cm}
 \caption{Diagrams included in the 2PI effective action: a) figure 8 b) 
pippi c) basketball. Full/dashed lines are $\chi$/$\varphi$ propagators. 
The crosses denote the inflaton mean field.
 }
 \label{fig:diagrams}
 \end{center} 
 \end{figure}

The series of diagrams in $\Gamma_2$ can be truncated, with each set of 
diagrams corresponding to an approximation to the full theory. In a 
weak-coupling expansion we include (see Fig.\ \ref{fig:diagrams}),
 \bea
&&\hm  i\Gamma_{2}[\phi,G_{\varphi},G_{\chi}] = 
 \frac{\lpc}{4} \int_\C d^4x\, G_{\varphi}(x,x) G_{\chi}(x,x) 
\\  \nonumber &&\hm 
 - \frac{ig^4}{4} \int_\C d^4xd^4y\, \phi(x) G_{\varphi}(x,y) 
 G_{\chi}^{2}(x,y) \phi(y) 
-\frac{ig^4}{8} \int_\C d^4xd^4y\, G_{\varphi}^{2}(x,y) 
 G_{\chi}^{2}(x,y).
 \eea
 Diagram b) is formally $\sim g^4\phi^2$. We have ignored the three-loop 
diagram that is $\sim g^8\phi^4$. 
All $g^2\phi^2$ insertions are resummed already at the Hartree level:
 therefore the truncation we use here formally amounts to assuming that 
$g^2\phi$ is small, but with no constraints on $g^2\phi^2$. Convergence 
properties of the 2PI coupling expansion were briefly discussed in Ref.\ 
\cite{Arrizabalaga:2005tf}, and for the $1/N$ expansion in the $O(N)$ 
model in Refs.\ \cite{Aarts:2001yn,Aarts:2006cv}.

 The equations of motion result from taking the functional derivatives,
 \bea
 \frac{\delta
\Gamma\left[\phi,G_{\varphi},G_{\chi}\right]}{\delta\phi(x)}=0,
  \;\;\;\;\;\;\;\;
 \frac{\delta \Gamma\left[\phi,G_{\varphi},G_{\chi}\right]}{\delta
G_{\chi}(x,y)}=0,
  \;\;\;\;\;\;\;\;
 \frac{\delta \Gamma\left[\phi,G_{\varphi},G_{\chi}\right]}{\delta
G_{\varphi}(x,y)}=0.
 \eea
 After going to momentum space and writing the contour propagators and 
self energies in terms of the statistical and spectral components, 
 \be
\label{eq:statspec}
 G_\chi(t,t';\kv) = F_\chi(t,t';\kv) - 
\frac{i}{2}\,{\rm sign}_{\cal C}(t,t') \rho_\chi(t,t';\kv),
 \ee
we find the standard equations 
\cite{Aarts:2002dj,Aarts:2001qa,Berges:2001fi}
 \bea
 \label{eq:2PIeom}
 \left[\partial_{t}^{2}+{\bf k}^{2}+M^{2}_{\varphi/\chi}(t)\right]
\rho_{\varphi/\chi}(t,t';\veck) =&&\hm -\int_{t'}^{t}dt''\,
\Sigma_{\varphi/\chi,\rho}(t,t'';\veck) \rho_{\varphi/\chi}(t'',t';\veck),
\nonumber \\
{}&&{} \\
 \left[\partial^{2}_{t} + {\bf k}^{2} + M^{2}_{\varphi/\chi}(t)\right]
F_{\varphi/\chi}(t,t';\veck) =&&\hm -\int_{0}^{t}dt''\,
\Sigma_{\varphi/\chi,\rho}(t,t'';\veck) F_{\varphi/\chi}(t'',t';\veck)
 \nonumber\\
 &&\hm +\int_{0}^{t'}dt''\, \Sigma_{\varphi/\chi,F}(t,t'';\veck)
\rho_{\varphi/\chi}(t'',t';\veck),
 \nonumber
 \eea
 with
 \bea
 \nn
 M_{\varphi}^{2}(t) = &&\hm m_{\varphi}^{2}+\lpc \int_\kv 
F_{\chi}(t,t;\kv),\\
 M_{\chi}^{2}(t) = &&\hm m_{\chi}^{2} + \lpc \left[ \phi^{2}(t) + 
\int_\kv F_{\varphi}(t,t;\kv) \right].
 \label{eq:hartmass}
 \eea
 The nonlocal self-energies originate from diagrams b) and c) and are 
easiest written in real space. They are
 \bea
 \label{eq:nonlocfirst}
 \Sigma_{\chi/\varphi}(x,y) = &&\hm \Sigma_{\chi/\varphi}^{(b)}(x,y) +
\Sigma_{\chi/\varphi}^{(c)}(x,y),
 \eea 
 where
 \bea
 \Sigma_{\chi/\varphi,\rho}^{(b)}(x,y) = &&\hm -2g^4\phi(x) \left[
F_{\chi}(x,y) \rho_{\varphi/\chi}(x,y) + F_{\varphi/\chi}(x,y)
\rho_{\chi}(x,y)\right] \phi(y), \nonumber\\
 \Sigma_{\chi/\varphi,F}^{(b)}(x,y) = &&\hm -2g^4\phi(x) \left[
F_{\chi}(x,y) F_{\varphi/\chi}(x,y) - \rho_{\varphi/\chi}(x,y)
\rho_{\chi}(x,y)/4 \right] \phi(y),
 \nonumber\\
&&
\eea
and
\bea
 \Sigma_{\chi/\varphi,\rho}^{(c)}(x,y) = 
 &&\hm -2g^4 \left[ F^{2}_{\varphi/\chi}(x,y) -
\rho^{2}_{\varphi/\chi}(x,y)/4 \right] \rho_{\chi/\varphi}(x,y),
 \nonumber \\
 &&\hm
 -4g^4 F_{\varphi/\chi}(x,y) \rho_{\varphi/\chi}(x,y)
F_{\chi/\varphi}(x,y)
 \nonumber\\
 \Sigma_{\varphi/\chi,F}^{(c)}(x,y) = &&\hm
 -2g^4 \left[ F^{2}_{\chi/\varphi}(x,y) - \rho^{2}_{\chi/\varphi}(x,y)/4
\right] F_{\varphi/\chi}(x,y)
 \nonumber\\
 &&\hm + g^4 F_{\chi/\varphi}(x,y) \rho_{\chi/\varphi}(x,y)
\rho_{\varphi/\chi}(x,y),
 \label{eq:nonloclast}	
 \eea
 The mean field equation is given by
 \bea
 \label{eq:mf}
 \left[ \partial^{2}_{t}+3H(t)\partial_{t}+M_{\phi}^{2}(t) \right] \phi(t) 
= -\int_0^t dt'\, \Sigma_{\phi}(t,t') \phi(t'),
 \eea 
 with $M_{\phi}^{2}(t) = M_{\varphi}^{2}(t)$, 
$H(t)$ determined by Eq.\ (\ref{eq:H}), and 
 \bea
 \Sigma_{\phi}(t,t') = \Sigma_{\varphi,\rho}^{(c)}(t,t';\veck=\vecnul).
 \eea

% SUBSECTION: NUMERICAL ANALYSIS

\section{Numerical analysis\label{sec:numerics}}
\setcounter{equation}{0}

In order to solve the set of coupled integro-differential equations
presented in the previous sections, the system is discretized on a
space-time lattice with spatial lattice spacing $a_s$ and temporal
spacing $a_t$. The resulting discretized 2PI equations are solved
numerically, see Refs.\
\cite{Berges:2001fi,Berges:2004yj,Arrizabalaga:2005tf} and references
therein for details. The initial density matrix is taken to be Gaussian,
such that only the one- and two-point functions need to be initialized.
For the one-point functions we take $\dot{\phi}(0)=0$ but a large
initial $\phi(0)=\phi_{0}=10 M_{\rm pl}$. For the two-point functions we
use
 \bea
 G_{\chi/\varphi}(t,t';\veck)\bigg|_{t=t'=0} = &&\hm 
 \frac{n_{\bf k}^{\chi/\varphi,0}+1/2}{\omega_{\bf k}^{\chi/\varphi,0}},
 \\
 \partial_t\partial_{t'} G_{\chi/\varphi}(t,t';\veck)\bigg|_{t=t'=0} = &&\hm
 \left(n_{\bf k}^{\chi/\varphi,0}+1/2\right)\omega_{\bf k}^{\chi/\varphi,0},
 \\
 \label{eq:dtG}
 \partial_t G_{\chi/\varphi}(t,t';\veck)\bigg|_{t=t'=0} = &&\hm 0,
 \eea
 where the initial mode energy is 
 \be
 \omega_{\bf k}^{\chi/\varphi,0} = \sqrt{\kv^2+ M_{\chi/\varphi}^2(0)},
 \ee
 and the initial particle number is
 \be 
 n_{\bf k}^{\chi/\varphi,0} = \left(\exp\left[
\omega_\kv^{\chi/\varphi,0}/T_{\chi/\varphi}\right] - 1 \right)^{-1}.
 \ee
 For the lattice discretized formulation, an ${\cal 
O}(a_t)$ improvement term is added to the RHS of Eq.\  (\ref{eq:dtG}), 
\be
\frac{1}{a_t}\left[ G_{\chi/\varphi}(a_t,0;\veck) - 
G_{\chi/\varphi}(0,0;\veck) \right] = 
-\frac{a_t}{2} \left( n_{\bf k}^{\chi/\varphi,0}+1/2\right)
\omega_{\bf k}^{\chi/\varphi,0},
\ee
 which ensures that, in the Hartree approximation, evolution initialized 
at the Hartree fixed point \cite{Aarts:2000wi} remains at that fixed 
point, for finite $a_t$.
 The initial conditions then read
 \bea
G_{\chi/\varphi}(a_t,a_t;\veck) = &&\hm 
 G_{\chi/\varphi}(0,0;\veck) = 
 \frac{n_{\bf k}^{\chi/\varphi,0}+1/2}{\omega_{\bf k}^{\chi/\varphi,0}},
 \\
G_{\chi/\varphi}(a_t,0;\veck) = &&\hm 
G_{\chi/\varphi}(0,0;\veck)
\left[ 1-\half \left( a_t \omega_\kv^{\chi/\varphi,0}\right)^2\right],
 \eea
 We introduce separate initial ``temperatures'' $T_{\chi,\varphi}$ for 
the $\chi$ and the $\varphi$ modes, so that several scenarios can be 
explored.
 The initial masses $M^{2}_{\chi/\varphi}(0)$ are obtained by solving the 
gap equations (\ref{eq:hartmass}) at $t=0$ with $\phi=\phi_{0}$. These 
equations are divergent and we use a simple local mass counterterm to take 
care of this. A much more sophisticated renormalisation scheme exists 
\cite{vanHees:2001ik,VanHees:2001pf,Blaizot:2003an,Berges:2005hc,Arrizabalaga:2006hj}, 
but this more straightforward approach has been seen to perform well 
\cite{Arrizabalaga:2005tf}. For the coupling constant we take $\lpc = 1$ 
throughout. The (renormalized) masses at zero temperature are 
$m_{\chi}/M_{\rm pl} = 0.1$ and $m_{\varphi}/M_{\rm pl} = 
1$.
 Finally, 
the spatial and 
temporal lattice spacings are taken as $a_sM_{\rm pl}=1$ and 
$a_t/a_s=0.01$. The spatial lattice discretisation is fairly coarse, 
especially at early times when $\phi(t)$ is still large. The time 
discretisation is well under control throughout.

 We compare a range of approximations described above and in the Appendix:
 \begin{enumerate}
 \item[1)] free: the mean field equation is solved without back-reaction, 
the $\chi$ propagator is solved in this background (Sec.\ \ref{sec:model}).
 \item[2)] perturbative: solution of the perturbatively expanded 
mean-field 
equation, the $\chi$ propagator is reconstructed in the 
perturbative coupling expansion (Appendix \ref{eq:perturbative}).
 \item[3)] Hartree: solution of the 2PI equations keeping diagram a) 
(Appendix \ref{sec:hartree}). 
 \item[4)] no pippi: solution to the 2PI equations keeping diagrams a) and 
c).
 \item[5)] full: solution of the 2PI equations keeping diagrams a), b) and 
c).
 \end{enumerate}

 Approximation 4) is used since it mimics the case considered in Ref.\ 
\cite{Berera:2004kc}, where the mean field equation includes the 
(perturbatively expanded) Hartree term, but with a dressed $\chi$ 
propagator. Including the basketball diagram c) amounts to dressing the 
$\chi$ and $\varphi$ propagators with the sunset self-energy; this affects 
the mean field only via the equal-time $\chi$ propagator. Comparing 4) and 
5) also gives the possibility to assess whether a given effect originates 
from diagram b) or c).

 We now explore different scenarios and vary the initial ``temperatures'' 
$T_{\chi/\varphi}$. Explicitly, we take vacuum initial conditions 
($T_{\chi/\varphi}=0$), partly thermal conditions ($T_{\chi/\varphi}\neq 
0$, $T_{\varphi/\chi}= 0$), and thermal conditions 
($T_{\chi/\varphi}\neq 0$).

% SUBSECTION: VACUUM INITIAL CONDITIONS

\subsection{Vacuum initial conditions\label{sec:Vacuumresults}}

\begin{figure}
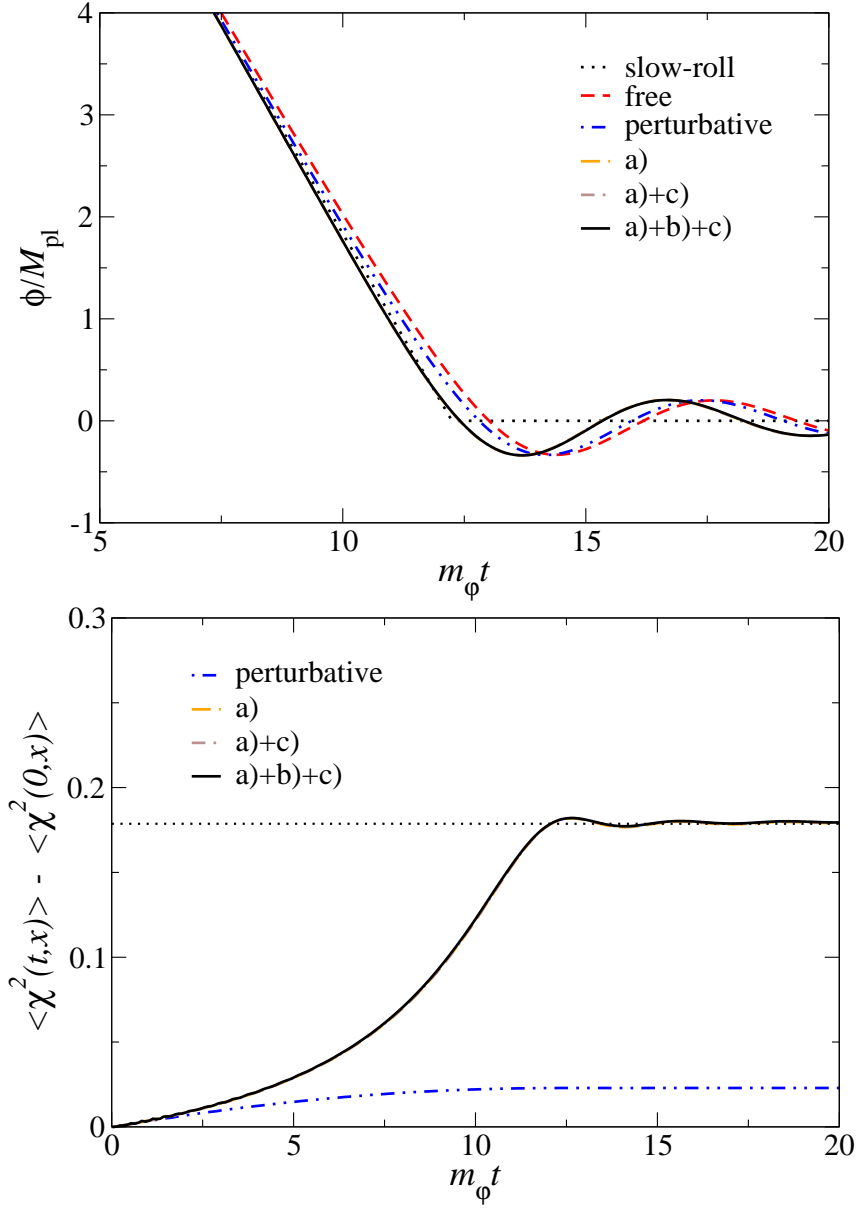

\begin{center}
\epsfig{file=MFT00.eps,height=8cm}
\epsfig{file=GXT00.eps,height=8cm}
 \caption{Evolution of the mean field (top) and the equal-time 
$\bra\chi^2\ket$ correlator (bottom) in the case of vacuum initial 
conditions, $T_{\chi} = T_{\varphi}=0$. The various approximations are 
slow-roll, free, perturbative, diagram a) (Hartree approximation), 
diagrams a) and c) (no pippi), and diagrams a), b) and c) (full) of Fig.\ 
\ref{fig:diagrams}. The evolution in the latter three cannot be 
distinguished. In the bottom figure the dotted line indicates the 
asymptotic result in the adiabatic approximation.
 }
 \label{fig:T0T0}
\end{center}
\end{figure}

We initialise the system with vacuum propagators in both the $\chi$ and 
the $\varphi$ sectors, $T_{\chi}=T_{\varphi}=0$. This is the relevant 
initial condition if we expect inflation to have strongly diluted all 
matter in the Universe, and all memory of the initial conditions have been 
lost. In order to heat the Universe, we require reheating (after 
inflation) or dissipation effects (during inflation) to create particles.

In Fig.\ \ref{fig:T0T0} (top) the mean field evolution in time is
presented, starting in all cases from $\phi_{0}/M_{\rm pl}=10$ with
vacuum propagators. Note that the plot only shows the evolution 
when $\phi(t)/M_{\rm pl} <4$ (or $m_\varphi t\gtrsim 7$); for larger 
values of $\phi$ the slow-roll approximation is very accurate.
 The evolution in various approximations is shown:
slow-roll, free, and perturbative, and the 2PI approximations
corresponding to the diagrams of Fig.\ \ref{fig:diagrams}: diagram a)
(Hartree approximation), diagrams a) and c) (no pippi), and diagrams a),
b) and c) (full).
 The slow-roll and free result agree initially, but start to deviate 
somewhat when $m_{\varphi}t\simeq 7$. Slow-roll ends around 
$\phi\simeq\sqrt{2}M_{\rm pl}$, or $m_{\varphi}t\simeq 10$.
 When including back-reaction, the three approximations a), a)+c), and 
a)+b)+c) are practically indistinguishable, but they differ from the free 
case. This is due to the time-dependent effective mass $M^{2}_{\phi}= 
m_\varphi^2+ g^2\bra \chi^2(t,\xv)\ket$, which increases in time and 
causes the mean field to roll down faster. The time-dependent part of the 
mean field mass is shown in Fig.\ \ref{fig:T0T0} (bottom) in the form of 
the 
subtracted equal-time $\bra\chi^2\ket$ correlator, $\bra \chi^2(t,\xv) 
\ket - \bra \chi^2(0,\xv) \ket$. Here, all approximations behave in the 
same way, and approach the result predicted by the adiabatic 
approximation, indicated by the horizontal dotted line. The odd one out is 
the perturbative approximation, which can only be trusted for times 
shorter than, say $m_{\varphi}t\simeq 2$. This also explains why the mean 
field evolution in the perturbative approximation is much closer to the 
free one; the back-reaction on the mean field is too small.

We find that the evolution is so slow that little departure from 
adiabaticity is observed and very few particles are created. It is 
therefore perhaps not surprising that no dissipative effects are seen, and 
that the Hartree approximation catches all the main features.

% SUBSECTION: FINITE T CHI

\subsection{Thermal initial conditions for the $\chi$ field
\label{sec:finiteTX}}

\begin{figure}[t]
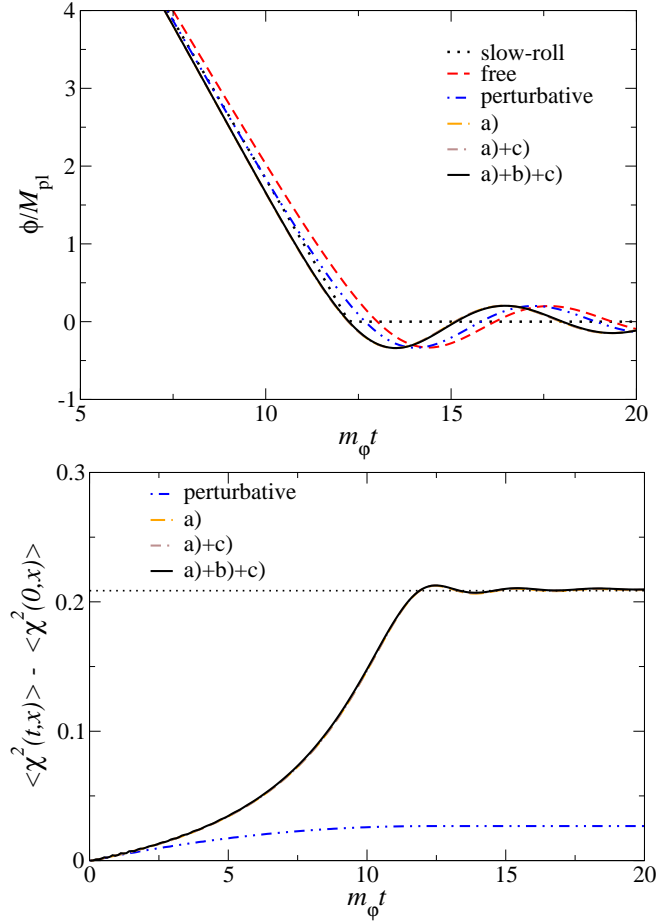
 \begin{center} 
\epsfig{file=MFT40.eps,height=6.1cm} 
\epsfig{file=GXT40.eps,height=6.1cm} 
 \caption{As in Fig.\ \ref{fig:T0T0}, for partly thermal initial 
conditions, $T_{\chi}/M_{\rm pl}=4$, $T_{\varphi}=0$.
 }
\label{fig:T4T0}
\end{center}
\end{figure}

We now consider the scenario where the $\chi$ propagator is initially at 
finite temperature, $T_{\chi}/M_{\rm pl}=4$, but the $\varphi$ propagator 
remains in vacuum, $T_{\varphi}=0$. This mimics the case when particle 
creation has been going on for a while in the $\chi$ propagator, but has 
not yet trickled down into the $\varphi$ propagator. In this context, the 
$\varphi$ propagator modes can be thought of as light particle degrees of 
freedom into which the heavy $\chi$ excitations, with $M_{\chi}\sim 
g\phi(t)$, may decay. This is motivated by the scenario where additional 
species are coupled to the $\chi$ sector 
\cite{Berera:2002sp,Aarts:2007qu}. If dissipative effects are important, 
these should show up via the nonlocal diagrams in the propagator and the 
mean field equation of motion.

In an expanding background, the particles would be redshifted away during 
inflation, unless sourced by particle creation from the inflaton rolling. 
By ignoring the expansion in the propagator equation of motion, we keep 
the initial and any subsequently produced particles, favouring dissipative 
effects. We note, that as the $\chi$ particles are very heavy initially, 
the initial particle numbers are very small, $n_\kv^\chi \simeq 
1/[\exp{(M_{\chi}/T_{\chi})}-1] \simeq 0.09$.

In Fig.\ \ref{fig:T4T0} we show the mean field evolution (top) and
$\bra\chi^2\ket$ correlator (bottom) for this case. The results are very 
similar
to the ones discussed above, although the back-reaction is stronger due
to the $\chi$ heatbath, causing the mean field to roll down slightly
faster.

We find therefore that the particles present in the $\chi$ sector 
influence the evolution of the $\chi$ propagator and the mean field very 
little.\footnote{In Ref.\ \cite{Aarts:2007qu} the $\chi$ propagator was 
initialised in vacuum. We have repeated the simulations in that setup with 
thermal initial conditions ($T_{\chi}/M_{\rm pl}=4$) and found no effect 
of the decay channel as well.}

% SUBSECTION: FINITE T CHI

\subsection{Thermal initial conditions for the $\varphi$ field
\label{sec:finiteTP}}

In the next scenario we initialise the $\varphi$ propagator in thermal 
equilibrium, with $T_{\varphi}/M_{\rm pl}=4$, and keep the $\chi$ sector 
in vacuum, $T_{\chi}=0$.  This mimics the scenario of Ref.\ 
\cite{Berera:2002sp}, where integrating out the $\varphi$ degrees of 
freedom is expected to yield a finite width for the $\chi$ particles. In 
this case particle numbers are of order $n_\kv^\varphi \simeq 
1/[\exp{(M_{\varphi}/T_{\varphi})}-1]\simeq 3.5$, so a significant 
heatbath is present.
 As long as the $\chi$ mass is larger than the $\varphi$ mass, one may 
except the heatbath to provide effective damping and a thermal mass. If 
the coupling is strong enough, the $\chi$ sector may thermalise as well. 
At the end of inflation, when $M_{\chi}(t)<m_{\varphi}$, the $\varphi$ 
particles in the heatbath may potentially decay directly into $\chi$ 
excitations.

\begin{figure}[t]
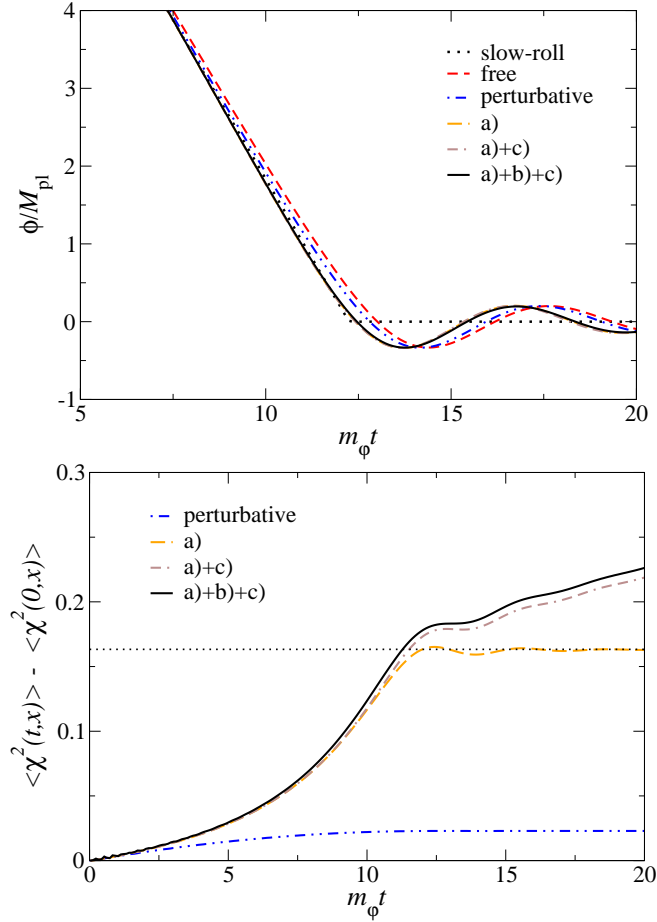

\begin{center}
\epsfig{file=MFT04.eps,height=6.1cm}
\epsfig{file=GXT04.eps,height=6.1cm}
 \caption{As in Fig.\ \ref{fig:T0T0}, for partly thermal initial 
conditions, $T_{\chi}=0$, $T_{\varphi}/M_{\rm pl}=4$.
 }
 \label{fig:T0T4}
\end{center}
\end{figure}

 The results for these initial conditions are shown in Fig.\ 
\ref{fig:T0T4}. The most striking effect is the change to the 
$\bra\chi^2\ket$ correlator. The presence of the $\varphi$ particles 
affects the evolution in those approximations that are sensitive to 
scattering in the heatbath, i.e.\ the approximations that contain the 
nonlocal diagrams b) (pippi) and c) (basketball). The Hartree 
approximation does not incorporate decay or scattering and cannot be 
applied to this part of the evolution. The fact that both approximations 
a)+c) and a)+b)+c) show the growing $\bra\chi^2\ket$ correlator indicates 
that it is mostly due to the presence of the basketball diagram. We note 
that the effect is largest after slow roll ends, but is visible during the 
latter part of the slow-roll regime as well.
 The mean field evolution is very close to the one with vacuum initial 
conditions, indicating that the $\varphi$ heatbath does not significantly 
backreact on the mean field evolution using the $\chi$ modes as 
intermediaries.
 It should be noted that the perturbative approximation does not know 
about the $\varphi$ propagator, and so is identical to the case with 
vacuum initial conditions.

% SUBSECTION FINITE T CHI AND VARPHI

\subsection{Thermal initial conditions for both fields
\label{sec:finiteTXTP}}

\begin{figure}[t]
\begin{center}
\epsfig{file=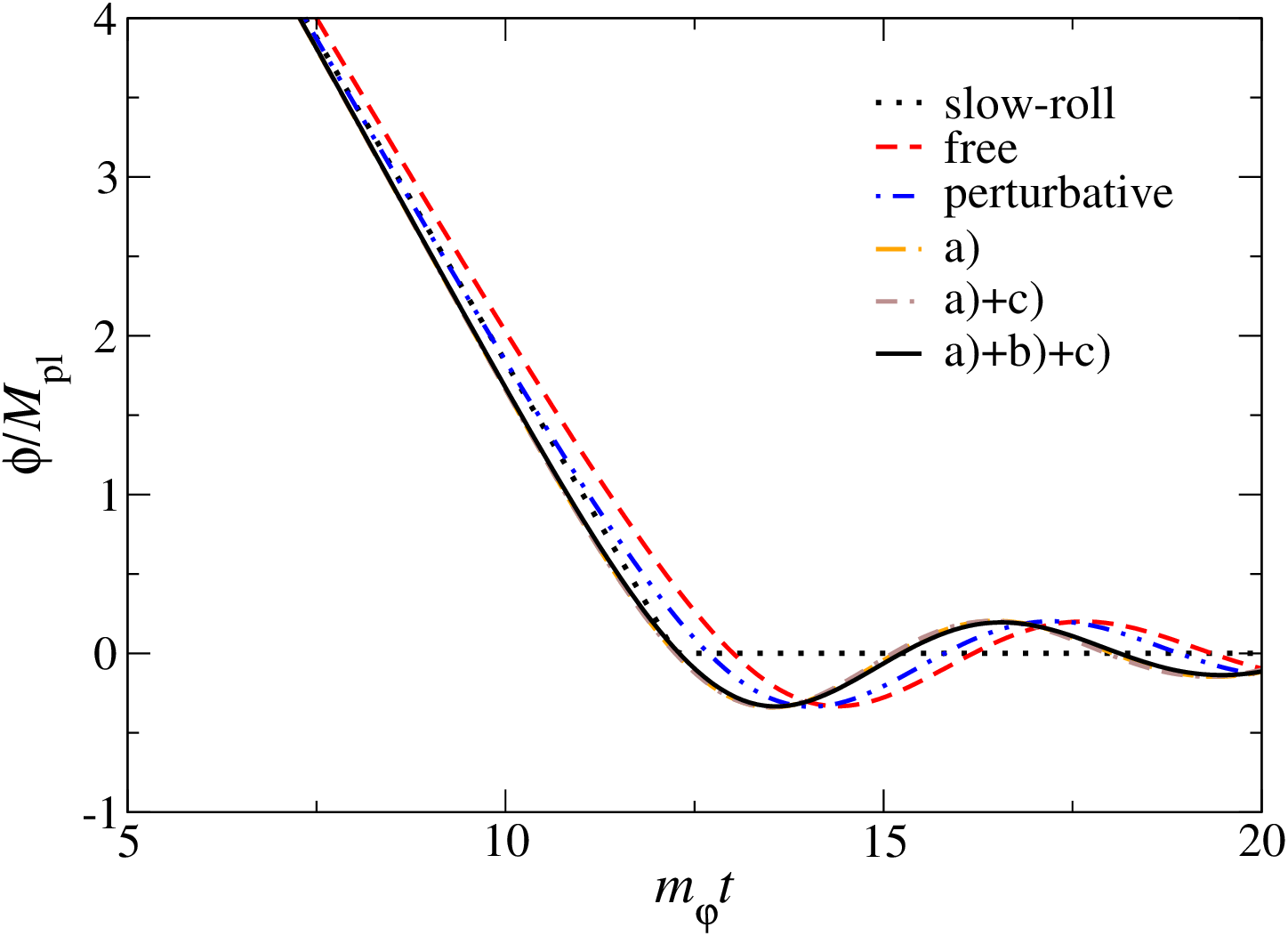,height=6.1cm}
\epsfig{file=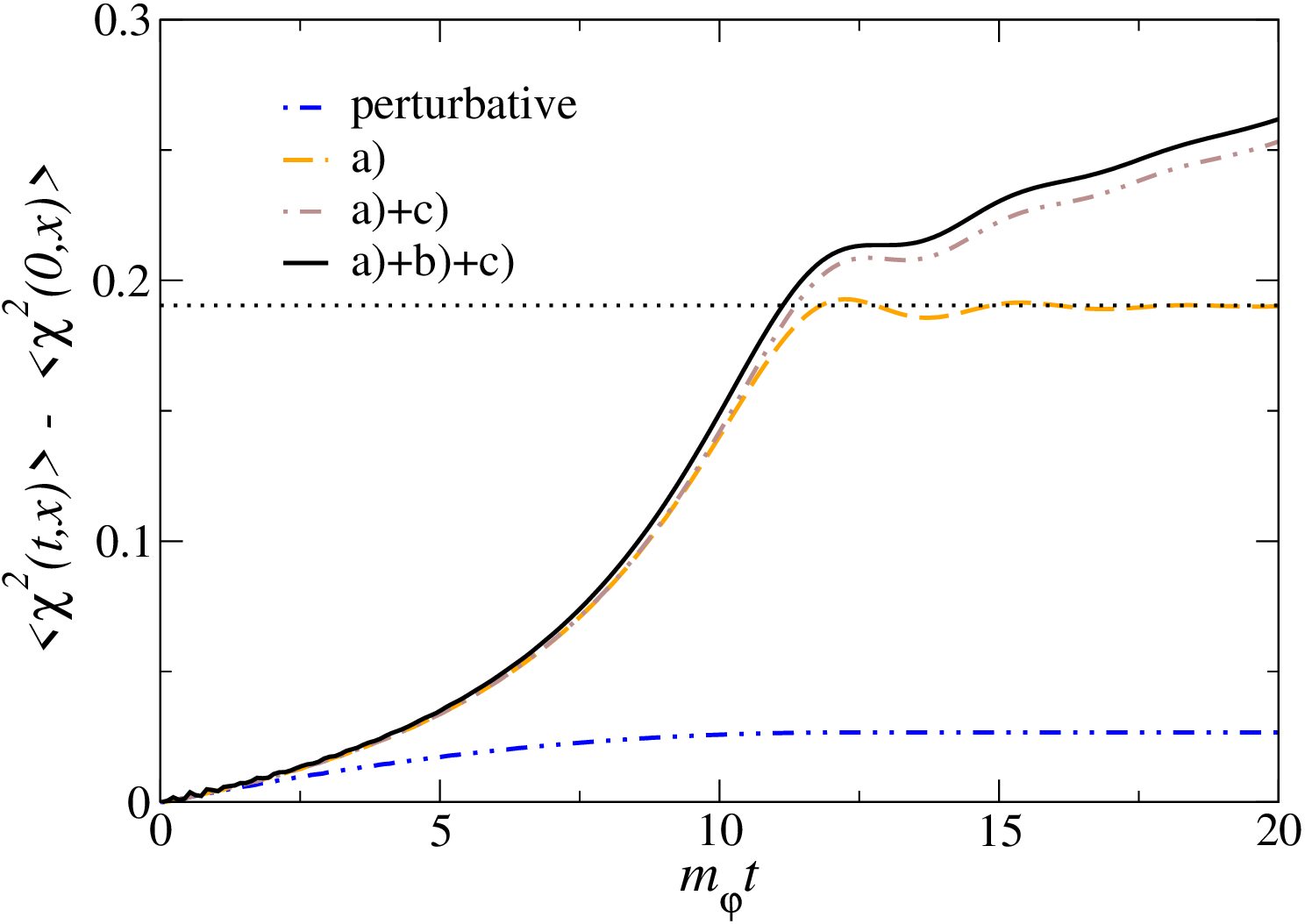,height=6.1cm}
 \caption{
As in Fig.\ \ref{fig:T0T0}, for thermal initial 
conditions, $T_{\chi}/M_{\rm pl}=T_{\varphi}/M_{\rm pl}=4$.
 }
 \label{fig:T4T4}
\end{center}
\end{figure}

As the final case, both fields are initialized in thermal equilibrium, 
with $T_{\chi}/M_{\rm pl}=T_{\varphi}/M_{\rm pl}=4$. This should further 
enhance the effects of the nonlocal diagrams and mimics the case when 
interactions are strong enough that the whole system is in equilibrium.

The results are shown in Fig.\ \ref{fig:T4T4} and are very similar to the 
case just discussed above. The heatbath of $\varphi$ particles affects 
the $\chi$ modes mostly after the end of slow roll, where energy transfer 
from the $\varphi$ to $\chi$ sector is effective. On the other hand, the 
heatbath of $\chi$ particles makes the effective mean field mass larger 
than in vacuum, which causes $\phi$ to roll down slightly faster. The 
nonlocal diagrams b) and c) become essential towards the end of inflation.

For completeness, we note that full equilibration and thermalization
takes place on much longer time scales. This has been studied in detail
using the loop expansion we used here in Ref.\ \cite{Berges:2000ur,Arrizabalaga:2005tf,Lindner:2005kv}. Again the
inclusion of diagrams b) and c) are essential for this.

 \section{Conclusions}
 \label{sec:conclusions}
 \setcounter{equation}{0}

 The main goal of this work was to apply techniques of out-of-equilibrium 
quantum field theory to the problem of interacting quantum fields during 
and after a slow-roll regime. Although this topic has a long history, a 
number of 
assumptions, approximations and ans\"atze are usually invoked along the 
way. Here we have employed a systematic formulation based on the loop 
expansion of the 2PI effective action. We considered a model where the 
 slow-rolling field $\varphi$ is coupled to a second scalar field $\chi$ 
via a $\varphi^2\chi^2$ interaction.
 In order to separate the question of dissipation and particle creation 
during slow-roll from the dilution in rapidly expanding spacetimes  
relevant for inflation, we have treated the quantum modes in Minkowski 
space but preserved the Hubble friction term in the mean field equation 
of motion. In this setup the mean field rolls slowly at the early stages, 
after which a smooth transition to the reheating phase is made. It also 
allowed us to bypass technical issues regarding the numerical solution and 
renormalisation in expanding space-times.

 We have studied several scenarios, corresponding to vacuum and thermal
initial conditions in the $\varphi$ and/or $\chi$ sector.
 The feature dominating the dynamics arises from the interplay between the 
mean field and the equal-time $\bra\chi^2\ket$ correlator. During the 
slow-roll regime, the $\bra\chi^2\ket$ correlator increases significantly 
which results in an increasing effective mean field mass.
  This effect is not important during the slow-roll stage, but eventually 
the increasing mean field mass drives the evolution away from slow-roll, 
such that this regime ends earlier than in the absence of interactions.
 We found that this is the main source of back-reaction on the mean field. 
It is already included at the level of the Hartree approximation and is 
well reproduced by the adiabatic approximation.
 The presence of the heatbaths affects the later stages of the evolution 
in various ways.
 A heatbath of $\chi$ particles mainly speeds up the end of slow roll, 
since it results in a larger effective mean field mass.
 A heatbath of $\varphi$ particles is important for the evolution of the 
$\chi$ modes, due to the possibility of creating $\chi$ particles from 
$\varphi$ particles. This effect is important mostly at the end of and 
after slow roll. In this case it is essential to include nonlocal diagrams 
beyond the mean field/Hartree approximation, which fails to capture this 
effect.
 In all scenarios the perturbative approximation was found to not perform 
very well and to be at best qualitatively valid for very early times, 
even in the slow-roll regime.

Particle creation and thermal back-reaction effects are crucial in warm 
inflation. We believe that the scenario would benefit from being recast in 
the 2PI formalism, since in that case no quasi-particle or equilibrium 
assumptions would have to be imposed on the propagators. Also, the 2PI 
expansion systematically includes all the scattering and dissipation 
effects relied upon in warm inflation. For the parameter values used in 
this work, we did not reach a warm inflation regime.

2PI resummed equations of motion have been successfully applied to 
reheating dynamics \cite{Berges:2002cz,Arrizabalaga:2004iw}. There is 
still the issue of the numerical implementation in expanding backgrounds, 
addressed in Refs.\ 
\cite{Boyanovsky:1993xf,Stephens:1998sm,Rajantie:2006gy}. Clearly, a 
complete description of inflationary dynamics requires this to be 
resolved.

% ACKNOWLEDGEMENTS

\vspace*{0.5cm}
\noindent
{\bf Acknowledgements.}
 We thank Jan Smit, Arjun Berera, Rudnei Ramos and Ian Moss for 
discussion. G.A.\ is supported by a PPARC Advanced Fellowship. A.T. is 
supported by the PPARC SPG ``Classical Lattice Field Theory'' and by the
Academy of Finland Grant 114371. This work 
was partly conducted on the COSMOS Altix 3700 supercomputer, funded by 
HEFCE and PPARC in cooperation with SGI/Intel, and on the Swansea Lattice 
Cluster, funded by PPARC and the Royal Society.

\appendix
\renewcommand{\theequation}{\Alph{section}.\arabic{equation}}

\section{Further approximations}

 In this Appendix, we present 
 a set of approaches for approximating the local correlator 
$G_\chi(x,x)=\bra\chi^2(x)\ket$. 
These are all well-known in the literature and are used in the main body 
of the paper for comparison.
 
% SUBSUBSECTION: WKB ANSATZ AND ADIABATICITY

\subsection{WKB ansatz, adiabaticity and particle creation\label{sec:WKB}}

We consider the evolution of the mode functions, determined by
 \be
 \label{eq:modeeq2}
 \ddot{f}_{\bf k}(t) + \om_\kv^2(t) f_{\bf k}(t)=0,
 \ee
with
 \be
 \om_\kv(t) = \left[\kv^2 + m_{\chi}^2 + \delta m^2(t) \right]^{\half},
\;\;\;\;\;\;
 \delta m(t) = g \left( \phi_0 - \sqrt{\frac{2}{3}} M_{\rm pl} m_\varphi t
\right).
 \ee
 As a first approximation it is instructive to introduce a WKB ansatz of the
form \cite{Kluger:1998bm}
 \be
 \label{eq:adia}
 f_\kv(t) = \frac{1}{\sqrt{2\Om_\kv(t)}} 
 \exp \left[ -i\int_0^tdt'\,\Om_\kv(t') \right].
 \ee
 It is assumed that $\Om_\kv(t) > 0$ for all $t$. Inserting this into Eq.
(\ref{eq:modeeq2}) one finds that
 \be
 \Om_\kv^2 = \om_\kv^2 -\frac{1}{2}\frac{\ddot \Om_\kv}{\Om_\kv} +
\frac{3}{4}\frac{\dot\Om_\kv^2}{\Om_\kv^2}.
 \ee
 In the adiabatic limit time derivatives are small and $\Om_\kv(t) \approx
\om_\kv(t)$, such that
 \be
 \label{eq:adiablim}
 \Om_\kv^2 = \om_\kv^2 -\frac{1}{2}\frac{\ddot \om_\kv}{\om_\kv} +
\frac{3}{4}\frac{\dot\om_\kv^2}{\om_\kv^2} + \ldots
 \ee
 Therefore the adiabatic approximation is valid provided
 \be
 \frac{\dot\om_\kv}{\om_\kv^2}\ll 1,
 \;\;\;\; \;\;\;\; \;\;\;\;
 \frac{\ddot\om_\kv}{\om_\kv^3}\ll 1.
 \ee
 The (adiabatic) particle number is then given by 
 \be
 n_\kv(t) = \frac{1+2n_\kv^0}{2\om_\kv(t)} \left[ |\dot f_\kv(t)|^2 + 
\om_\kv^2(t)
|f_\kv(t)|^2 \right] -\frac{1}{2} \simeq n_\kv^0 + 
{\cal O}\left(\dot\om_\kv^2/\om_\kv^4\right),
 \ee
 where $n_\kv^0$ is the initial $\chi$ particle number present in mode 
$\kv$. This is the usual adiabatic result: particle creation is controlled 
by the rate of change $\dot\om_\kv/\om_\kv^2$ \cite{Kluger:1998bm}.

 However, the absence (or suppression) of particle creation does not mean
that equal-time correlators are constant. For the equal-time correlator
 \be
 \bra \chi(t,\xv)\chi(t,\yv) \ket = G_\chi(t,t; \xv-\yv) 
 = \int_\kv  e^{i\kv\cdot(\xv-\yv)} G_\chi(t,t; \kv),
\ee
 we find from the expressions above that
 \be
 \label{eq:Xcorr2}
 G_{\chi}(t,t;\kv) = \left( 1+2n_\kv^0 \right) |f_\kv(t)|^2 \approx \left(
1+2n_\kv^0 \right) \frac{1}{2\om_\kv(t)}.
 \ee
 Therefore we find that this equal-time correlator grows in time, with a   
time dependence directly determined by the time dependence of the 
mean field, since $\om_\kv(t)\sim \phi(t)$ for small $k$. 
 Similarly, the (unrenormalized) energy density in the $\chi$ modes,
\be
 E_\chi(t) = \int_\kv \frac{\om_\kv(t)}{2} \left( 2n_\kv(t) + 1 \right)
\simeq \int_\kv \frac{\om_\kv(t)}{2} \left( 2n_\kv^0 + 1\right),
\ee 
 decreases in time as $\phi(t)$ rolls down. 

% SUBSUBSECTION: EXACT SOLUTION

\subsection{Exact solution\label{sec:exact}}

 In certain mean field backgrounds, the mode equation (\ref{eq:modeeq2}) can 
be solved in closed form. In Ref.\ \cite{Aarts:2007qu}, we considered 
exponential time dependence $\phi(t)\sim e^{-\gamma t/2}$ (or $\delta 
m^2(t)\sim e^{-\gamma t}$) and found the analytical solution in terms of 
Bessel functions.\footnote{This solution is also relevant for a 
slow-rolling mean field in a $\phi^{4}$ potential.} We now perform a similar 
calculation for a slow-rolling mean field in a $\phi^{2}$ potential, where 
$\phi(t)$ is given by Eq.\ (\ref{eq:SRphi}).

 It is useful to define
 \be
 \kappa \equiv \left(-\frac{d\delta m(t)}{dt}\right)^{\half} 
 = \left( g \sqrt{\frac{2}{3}} M_{\rm pl}m_\varphi \right)^\half,
 \ee
 and write
 \be
 \label{eq:analfirst}
 f_\kv(t) = \exp\left( i g\phi_0t + \frac{i}{2}\kappa^2 t^2 \right)
h_\kv(t).
 \ee
 We find that $h_\kv(t)$ satisfies
 \be
 \ddot h_\kv(t) + 2i \delta m(t) \dot h_\kv(t) + \left(
\om_\kv^2-i\kappa^2\right) h_\kv(t) = 0,
 \ee
 where $\om_\kv=\sqrt{\kv^2+m_\chi^2}$ [such that $\om_\kv^2(t) = 
\om_\kv^2 + \delta m^2(t)$]. Changing variables to
 \be
 x = \frac{\delta m(t)}{\sqrt{i}\kappa},
 \ee
 yields Hermite's equation,
 \be
 h_\kv''(x) - 2x h_\kv'(x) + 2\nu_\kv h_\kv(x) = 0,
 \ee
 where
 \be
 \nu_\kv = -\half \left( 1+i\frac{\om_\kv^2}{\kappa^2}\right).
 \ee
 The solution reads
 \be
 h_\kv(x) = A_\kv H_{\nu_\kv}(x) + B_\kv
\,{}_1F_1\left(-\frac{\nu_\kv}{2},\half,x^2\right),
 \ee
 where $H_\nu(x)$ is a Hermite function and 
${}_1F_1(\alpha,\beta,\gamma)$ is Kummer's confluent hypergeometric 
function \cite{AbraSteg}. The coefficients $A_\kv$ and $B_\kv$ are 
determined by the initial conditions, $f_\kv(0)=1/\sqrt{2\Omega_\kv}$ and 
$\dot f_\kv(0)= -i\Omega_\kv f_\kv(0)$, where $\Omega_\kv = 
\sqrt{\om_\kv^2+\delta m^2(0)}$. We find that
 \bea
 A_\kv = &&\hm \frac{1}{\sqrt{2\Omega_\kv} D_\kv}\left[ \alpha^{(22)}_\kv 
+ i\Omega_\kv \alpha^{(12)}_\kv \right], \\
 B_\kv = &&\hm \frac{-1}{\sqrt{2\Omega_\kv} D_\kv}\left[ 
\alpha^{(21)}_\kv + i\Omega_\kv \alpha^{(11)}_\kv \right],
 \eea
 where
 \bea
 \alpha^{(11)}_\kv = &&\hm H_{\nu_\kv}(x_0), \\
 \alpha^{(12)}_\kv = &&\hm
        {}_1F_1\left(-\frac{\nu_\kv}{2},\frac{1}{2},x_0^2\right), \\
 \alpha^{(21)}_\kv = &&\hm ig\phi_0 H_{\nu_\kv}(x_0) +
 2\nu_\kv \sqrt{i}\kappa H_{\nu_\kv-1}(x_0),
  \\
 \alpha^{(22)}_\kv = &&\hm
 ig\phi_0 \,{}_1F_1\left(-\frac{\nu_\kv}{2},\frac{1}{2},x_0^2\right) -
 2\nu_\kv \sqrt{i}\kappa x_0
\,{}_1F_1\left(1-\frac{\nu_\kv}{2},\frac{3}{2},x_0^2\right),
 \label{eq:anallast}
 \eea
 with $x_0=x(0)=\delta m(0)/\sqrt{i} \kappa$ and
$D_\kv = \alpha^{(11)}_\kv \alpha^{(22)}_\kv - \alpha^{(12)}_\kv 
\alpha^{(21)}_\kv$.

% SUBSECTION: HARTREE APPROXIMATION

\subsection{Hartree approximation\label{sec:hartree}}

 In the Hartree approximation, the connected three-point function is 
neglected and only one- and two-point functions are preserved (Gaussian 
approximation) \cite{Cooper:1996ii}.  In that case the back-reaction of 
the $\chi$ modes on the mean field manifests itself as a time-dependent mass 
and we find
 \be
 \label{eq:backreact}
 \ddot{\phi}(t) + 3H(t)\dot{\phi}(t) + M_{\phi}^{2}(t)\phi(t) = 0,
 \ee
 with
 \be
 M_\phi^2(t) = m_{\varphi}^{2} + \lpc \bra\chi^2(x)\ket 
 = m_{\varphi}^{2} + \lpc \int_\kv G_\chi(t,t;\kv).
 \ee
As is well known \cite{Cooper:1996ii}, the Hartree approximation as 
discussed here is the generalization to time-dependent systems of the 
standard approach based on the one-loop effective potential in 
equilibrium. In thermal equilibrium, after integrating out the $\chi$ 
field, one finds that
 \be
 V_{\rm 1 loop}(\phi) = \half T\sum_{n}\int_\kv 
\ln[\om_n^2+k^2+m_\chi^2+g^2\phi^2], 
 \ee
 where $\om_n=2\pi nT$ ($n\in \mathbb{Z}$) are the Matsubara frequencies.
One finds that 
 \be
 \frac{\partial V_{\rm 1 loop}(\phi)}{\partial \phi} = 
 g^2\bra\chi^2(x)\ket \phi,
 \ee
 where in equilibrium $\bra\chi^2(x)\ket$ is time-independent and, after 
performing the Matsubara sum, given by 
 \[
 \bra\chi^2(x)\ket = \int_\kv \frac{1+2n_B(\bar\om_\kv)}{2\bar\om_\kv},
 \;\;\;\;\;\;\;\; \;\;\;\;\;\;\;\;
 \bar\om_\kv=(\kv^2+m_\chi^2+g^2\phi^2)^\half.
 \]
 Here $n_B(\om)=1/(\exp(\om/T)-1)$ is the Bose distribution.

 Out of equilibrium, the evolution of the quantum fluctuations can be 
determined self-consistently in the Gaussian approximation in terms of the 
$\varphi$ and $\chi$ propagators as
\bea
i\left[\partial_{t}^{2} + \kv^2 + M_\varphi^2(t)\right] G_\varphi(t,t';\kv)
 =&&\hm \delta_{\mathcal{C}}(t-t'), \\
 i\left[\partial_t^2 + \kv^2 + M_\chi^2(t)\right] G_\chi(t,t';\kv) 
 = &&\hm \delta_{\mathcal{C}}(t-t'),
 \label{eq:harteom}
 \eea
 with the effective masses
 \bea
 M_\varphi^2(t) = &&\hm m_\varphi^2 + \lpc \int_\kv G_\chi(t,t;\kv),\\
 M_\chi^2(t) = &&\hm m_\chi^2 + \lpc \left[ \phi^2(t) 
 + \int_\kv G_\varphi(t,t;\kv)\right].
 \eea
The Hartree approximation is self-consistent and easily solvable numerically.

%% SUBSECTION: PERTURBATIVE HARTREE APPROXIMATION

\subsection{Perturbative approximation
\label{eq:perturbative}}

 \begin{figure}[t]
 \begin{center}
 \epsfig{file=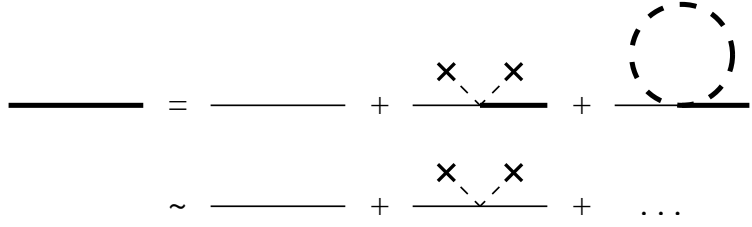,width=10cm}
 \caption{Graphical representation of the dynamics of the $\chi$ 
propagator (thick line) in the Hartree approximation. The thin line is the 
lowest order $\chi$ propagator, the thick dashed line the $\varphi$ 
propagator, and the crosses the mean field background. 
The second line shows the perturbative expansion discussed in 
Section \ref{eq:perturbative}.
 }
 \label{fig:hartree}
 \end{center}
 \end{figure}

We can also attempt to include the effect of the time-dependent mean field 
on the evolution of $G_\chi$ in a perturbative manner. This approach is 
indicated in Fig.~\ref{fig:hartree}: while in the Hartree approximation 
all $\phi^2$ insertions are summed nonperturbatively, in the perturbative 
setup only a single $\phi^2$ insertion is preserved.\footnote{This 
approximation is motivated by Refs.\ 
\cite{Gleiser:1993ea,Berera:1998gx,Berera:2004kc}.} To obtain this, we 
write
 \be
 \label{eq:pertexp}
 G_\chi(x,y) \equiv \bra T_{\cal C}\chi(x)\chi(y)\ket = 
 G_\chi^{(0)}(x,y) + \lpc G_\chi^{(1)}(x,y) + \mathcal{O}(g^4),
 \ee
 and solve Eq.\ (\ref{eq:harteom}) formally order by order in $\lpc$.  In
order to avoid large corrections due to the initial large expectation value
of the mean field, we write
 \be
 \label{eq:deltaphi} 
 \phi^2(x) = \phi^2(0) + \left[ \phi^2(x) - \phi^2(0) \right] 
 \equiv \phi_0^2 + \Delta\phi^2(x),
 \ee
 and treat $g^2\Delta\phi^2(x)$ as the perturbation. 
The expansion is therefore expected to be valid for early times only. 
We also ignore fluctuations of $\varphi$, since 
$G_\varphi(x,x)\ll \phi_0^2$ after renormalization. 
 In terms of the free inverse propagator, 
 \be
 G_0^{-1}(x,y) = i\left[ \square_x + m_\chi^2 + \lpc\phi_0^2 \right]
\delta_{\cal C}(x-y),
 \ee
we find the series of equations
 \bea
 \int_{\cal C} d^4z\, G_0^{-1}(x,z) G_\chi^{(0)}(z,y)
  =&&\hm \delta_{\mathcal{C}}(x-y), \\ 
 \int_{\cal C} d^4z\, G_0^{-1}(x,z) G_\chi^{(1)}(z,y)
  =&&\hm -i \Delta\phi^2(x) G_\chi^{(0)}(x,y), 
 \eea
 etc. 
The formal solution of the second equation can be written as (see Fig.\ 
\ref{fig:hartree})
 \be
 G_\chi^{(1)}(x,y) = -i \int_{\cal C} d^4z\, G_\chi^{(0)}(x,z) \Delta\phi^2(z)
G_\chi^{(0)}(z,y).
 \ee
 To first order in this expansion the mean field equation of motion then
reads \cite{Berera:2004kc}
 \be
 \label{eq:perturbeom}
 \left[\partial_t^2 + 3H(t)\partial_t + \bar M_\varphi^{2}(t) \right] 
\phi(t) =
 i g^4 \int_{\cal C} d^4z\, G_\chi^{(0)}(x,z) \Delta\phi^2(z) G_\chi^{(0)}(z,x)
 \phi(x),
 \ee
 with
 \be
  \bar M_\varphi^{2}(t) = m_\varphi^2 + \lpc G_\chi^{(0)}(x,x).
 \ee
 To make the time dependence explicit, we write the propagator along the 
Schwinger-Keldysh contour in terms of the statistical and spectral 
components (\ref{eq:statspec}) and find
 \be
 \label{eq:perturbeom2}
 \left[\partial_{t}^{2}+3H(t)\partial_{t}+\bar M_\phi^{2}(t) \right] 
\phi(t)= 
 \int_0^t dt'\, K(t,t') \Delta\phi^2(t')  \phi(t),
 \ee
 with the kernel
 \be
 \label{eq:pertkernel}
 K(t,t') = 2 g^4\int d^{3}x\, F_\chi^{(0)}(t,t'; \xv) 
 \rho_\chi^{(0)}(t,t'; \xv).
 \ee
 In order to solve  Eq.\ (\ref{eq:perturbeom2}) numerically, we need to 
choose a propagator for the memory kernel (\ref{eq:pertkernel}). We take the 
simplest perturbative ansatz
 \bea
 F_\chi^{(0)}(t,t';\kv) = &&\hm \frac{n_\kv^0 + 1/2}{\tilde{\omega}_\kv} 
 \cos[\tilde{\omega}_\kv(t-t')],\\
 \rho_\chi^{(0)}(t,t';\kv) = &&\hm 
 \frac{\sin[\tilde{\omega}_\kv(t-t')]}{\tilde{\omega}_\kv},
 \eea
 depending on the initial energies $\tilde{\omega}_{\bf k} = 
(\kv^2+m_\chi^2+\lpc\phi_0^2)^{\half}$.
 One may think of more elaborate ans\"atze for these propagators and 
include e.g.\ more thermal effects as well as the effect of scattering and 
decay processes \cite{Morikawa:1986rp,Berera:2004kc}. However, we prefer 
to discuss this in a systematic approach and employ the 2PI effective 
action to consider extensions beyond the mean field approximation.

 Finally, one can consider further approximations to these equations, 
using a derivative expansion 
\cite{Gleiser:1993ea,Berera:1998gx,Berera:2004kc}. However, we will not do 
so here.

%***************************************************************

\end{document}